\documentclass{llncs}
\usepackage{epsf}

\newcommand{\spin}{{\sc Spin}}
\newcommand{\promela}{{\sc Promela}}

\begin{document}

\sloppy

\pagestyle{headings}  
\pagestyle{plain}     

\addtocmark{SPINning Parallel Systems Software}


\title{SPINning Parallel Systems Software\thanks{
This work was supported by the Mathematical, Information, and
Computational Sciences Division subprogram of the Office of Advanced
Scientific Computing Research, U.S.  Department of Energy, under
Contract W-31-109-Eng-38.}}

\titlerunning{SPINning Parallel Systems Software}

\author{Olga Shumsky Matlin \and Ewing Lusk  \and William McCune}

\authorrunning{Matlin, Lusk, McCune}

\tocauthor{Olga Shumsky Matlin (Argonne National Laboratory),
  Ewing Lusk (Argonne National Laboratory),
  William McCune (Argonne National Laboratory)}

\institute{Mathematics and Computer Science Division\\
Argonne National Laboratory\\
Argonne, IL 60439, USA\\
\email{\{matlin,lusk,mccune\}@mcs.anl.gov}}

\maketitle              

\begin{abstract}

We describe our experiences in using \spin\ to verify parts of the
Multi Purpose Daemon (MPD) parallel process management system.  MPD is
a distributed collection of processes connected by Unix network
sockets.  MPD is dynamic: processes and connections among them are
created and destroyed as MPD is initialized, runs user processes,
recovers from faults, and terminates.  This dynamic nature is easily
expressible in the \spin/\promela\ framework but poses performance and
scalability challenges.  We present here the results of expressing
some of the parallel algorithms of MPD and executing both simulation
and verification runs with \spin.

\end{abstract}

\section{Introduction}
\label{sec:introduction}

Reasoning about parallel programs is surprisingly difficult.  Even
small parallel programs are difficult to write correctly, and an
incorrect parallel program is equally difficult to debug.  In our
experience writing the Multi Purpose Daemon (MPD), a parallel system
program described below, this characterization has been borne out:
despite MPD's small size and apparent simplicity, errors have impeded
progress toward code in which we have complete confidence.  (Of
course, another possible explanation is that as parallel programmers
we are inept; we reject this hypothesis because it is impossible to
verify rigorously.)  

Such a situation motivates us to explore program verification
techniques. Since our programs are small and our algorithms simple
(when viewed from the perspective of a single process), we hope that
program verification software will be able to handle our problem;
since the difficulty of reasoning about parallelism has shown us that
we really do need help with this problem, investing in verification is
worth the effort.   

MPD~\cite{bgl00:mpd:pvmmpi00,butler-lusk-gropp:mpd-parcomp} is a
process manager for parallel programs and is itself a parallel
program.  Its function is to start the processes of a parallel job in
a scalable way, manage input and output, deal with faults, and cause
jobs to terminate cleanly.  While the job is running, it may need to
provide services to the application, such as implementing a barrier or
assisting an application process in setting up communication with
another process in the job.  MPD is the sort of process manager needed
to run applications that use the standard
MPI~\cite{mpi-forum:journal,mpi-forum:mpi2-journal} library for
parallelism, although it is not MPI-specific.  MPD is distributed as
part of the portable and publicly available 
MPICH~\cite{gropp-lusk:mpich-www,gropp-lusk-doss-skjellum:mpich}
implementation of MPI. 

Our first attempt \cite{mpd-acl2} to use formal verification techniques to
ensure correctness of MPD algorithms was based on the ACL2
\cite{acl2:book,acl2} theorem prover.  While ACL2 provided a useful simulation
environment, formulating desired properties of and reasoning about models of
MPD algorithms proved difficult.  Our second approach, described here, employs
the model checker \spin~\cite{spin:book,spin:article}.
%
%
%
Our particular application is unusual in that the number
of entities and the topology of the communication network can change
over time.  We also need to model a larger number of entities than
many \spin\ applications do.  In Section~\ref{sec:results} we discuss
the challenges that these properties provide to the \spin\ system.

In Section~\ref{sec:approach} we describe MPD in more detail and
outline our method for modeling a distributed, dynamic set of Unix
processes in \promela.  In Section~\ref{sec:experiences} we describe 
our experiences with this approach, which we believe shows potential
benefits for the further development of MPD.  In
Section~\ref{sec:results} we present the concrete results of specific
verification experiments, and we conclude in Section~\ref{sec:plans}
with a summary of the current project status and our future plans.

\section{Approach}
\label{sec:approach}

To present what we want to verify and how we have gone about it, we
describe here a few salient features of the MPD system. More details
can be found in~\cite{bgl00:mpd:pvmmpi00}   
and~\cite{butler-lusk-gropp:mpd-parcomp}.

The MPD system consists of several types of processes.
The {\em daemons\/} are persistent (may run for weeks or months
at a time, starting many jobs), and there typically exists one daemon 
instance per host in a TCP-connected network.  {\em Manager\/}
processes are started by the daemons to control the application
processes ({\em clients\/}) of a single parallel job and provide most
of the MPD features.  The daemons are connected in a ring.  A {\em
  console\/} process is started by a user or another process to
connect to the daemon ring and give it a command, such as {\tt mpirun}
to start a user parallel job.  Separate managers for each user
process, started by the daemons, support individual process
environments for the user processes.  The managers also connect
themselves into a ring. 

Exactly how the daemons are started or connected is not important,
since the system provides a number of choices, and the process
need not be particularly fast.  A console command is started by the
user, either interactively or under the control of a batch scheduler.
The daemons spawn the managers, which use information given them by
the daemons to connect themselves into a ring, then spawn the clients.
The startup messages traverse the ring quickly, so most invocation of
new processes and connecting takes place in parallel, leading to fast
startup even for jobs involving hundreds of processes.
\begin{figure}[ht]
    \centerline{ \epsfxsize=4.5in \epsfbox{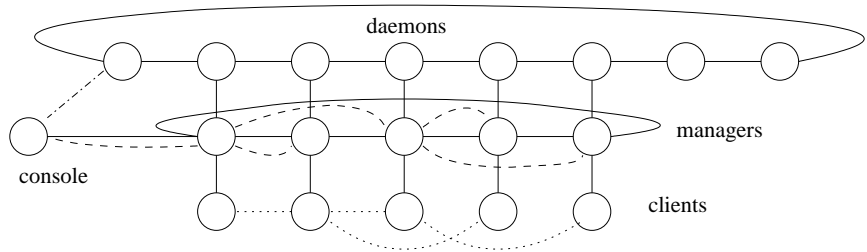} }
    \caption{Daemons with console process, managers, and clients}
    \label{fig:mpds-all}
\end{figure}
The situation is as shown in Figure~\ref{fig:mpds-all}, where the clients
may be application MPI processes.  The vertical solid lines represent
connections based on pipes; the remaining solid lines all represent
connections based on Unix sockets.  The dashed lines represent the
trees of connections for forwarding input and output, and the dotted
lines represent {\em potential\/} connections among the client
processes.  The dot-dashed line is the original connection from
console to local daemon on a Unix socket, which is replaced during
startup by the network connection to the first manager. 

An important feature of MPD for our purpose here is that the structure of the
code for each of the three process types (daemon, manager, console) is
essentially the same:  after initialization, the process enters an essentially
infinite loop, implemented by a Unix socket function \texttt{select},
which indicates which sockets have messages available.
That is, most of the time, the process is idle.  When a message
arrives on one of its sockets from one of the processes it is
connected to, it wakes up, parses the message, calls the appropriate
handler routine to process the message, and re-enters the idle state
by calling {\tt select} again.  The handler routine itself does a
small amount of processing, typically resulting in the creation of new
sockets or in the sending of messages on existing sockets.   The logic of
the distributed algorithms executed by the system as a whole,
therefore, is contained primarily in the handlers.  No individual
message, received by one process, results in much activity.  The
individual handlers are typically implemented in a few lines, or at
most a few tens of lines, of C code. 

\begin{figure}[t]
    \centerline{ \epsfxsize=2.0in \epsfbox{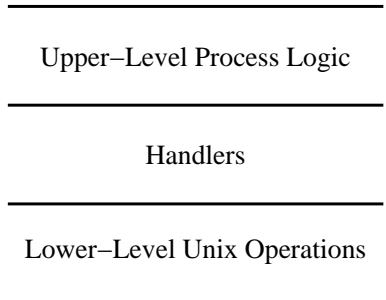} }
    \caption{Layers of code in MPD}
    \label{fig:layers}
\end{figure}

This structure allows us to treat the system as comprising three layers (see
Figure~\ref{fig:layers}).
The top tier corresponds to the upper-level logic of the process
(initialization, {\tt select} loop, parsing and dispatching of incoming
messages to handlers).  This is sequential logic that we have confidence in;
hence, although we must model this part of each process in some way in
\promela, the \promela\ model does not have to be faithful to the
algorithms in this layer.  The bottom layer corresponds to
well-understood Unix operations on sockets.  Again the code is
sequential and not of particular interest.  It is in the middle layer
(the handlers) that the interesting parallel algorithms are expressed
and the bugs appear.  

Verification of the algorithms executed by different components of the
MPD system is not the only goal of the project.   Equally important and
desirable is the ability to translate, possibly automatically,
a verified \promela\ model to executable C code while preserving the
verified properties of the model.  Only the middle layer of the model
has to be translated.  Thus, in our \promela\ model of each type
of process, we take considerable liberties with the top and bottom
layers, but we wish the Promela code for the middle layer (the handlers) to
be directly translatable into C or the scripting language Python.  Our
current implementation of MPD is in C.

\subsection{Modeling Components of the MPD System}

Components of the MPD system map naturally to \promela\ entities: a
{\tt proctype} is defined for each different MPD process type;
individual daemon, manager, console, and client processes correspond to 
active instances of the corresponding {\tt proctype}s; sockets map to
channels; and messages that are read and written over the sockets
correspond to messages traveling on the channels.  Our models of
individual process types preserve the three-layer structure
(Figure~\ref{fig:layers}) of executable MPD code for clarity,
readability, and modularity and also to facilitate the
translation of verified handler algorithms to executable code.

The top tier of the model typically contains the start-up logic of a
particular MPD algorithm followed by a mechanism for sampling and
processing the attached channels.  The middle tier corresponds to a
collection of handlers that define the set of message types permitted
for each type of communication link as well as the behavior of the
process in response to each message type.  The lowest tier consists of
a \promela-based implementation of the Unix socket primitives.  Note
that an MPD programmer relies on the predefined functionality of the
socket processing functions but does not implement the primitives.
Creating a \promela\ library of Unix socket primitives allows us to
(1) hide the details of the socket model from both the verification
and translation to the executable code, (2) interchange, if need be,
different models of sockets without changing the remainder of the
model, and (3) reuse the socket model in verification of independent
MPD algorithms.  

\subsection{Modeling Unix Sockets}

A Unix socket is an endpoint of a bidirectional communication path
between two processes.  In MPD, sockets are manipulated when a
connection between two processes is established or destroyed.
Correct operation of some MPD algorithms depends on correct allocation
and manipulation of sockets, while other MPD algorithms assume a
static system of processes and communication links between them.
Our objective is to model sockets efficiently to support verification of
both types of MPD algorithms.  

An MPD process runs on top of an operating system, which provides
(among other services) the implementation of sockets and their 
handling.  We must model the sockets, but we do so without creating a
model of the operating system.  In fact in our model, the operating
systems of the hosts on which MPD processes execute are combined into a
single implicit global operating system.  The only explicit
manifestation of the operating system is the socket descriptor
structure described below.  In the model, the functionality of the
operating system is hidden inside the socket library, and its tasks
are handled directly by the MPD processes.

\begin{figure}[t]
    \centerline{ \epsfxsize=4.5in \epsfbox{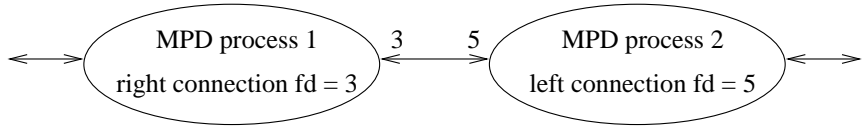} }
    \caption{Connected MPD processes}
    \label{fig:skt-example-a}
\end{figure}

\begin{figure}[t]
    \centerline{ \epsfxsize=3.0in \epsfbox{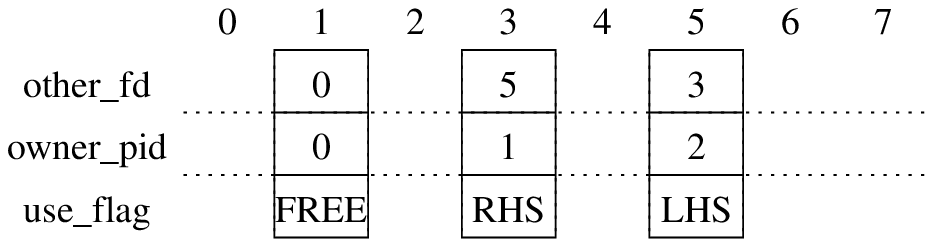} }
    \caption{Example state of the socket array}
    \label{fig:skt-example-b}
\end{figure}

A Unix socket is referenced by a file descriptor (fd) and represents
a buffer for reading and writing messages.  Our \promela\ model
of a socket consists of a channel and a three-part record that
describes how the particular socket should be used.  A model of an MPD
algorithm contains an array of channels and an array of socket
descriptor structures.  The first field of the socket descriptor
structure references the fd at the other endpoint of the connection. The
second field identifies a process that has an exclusive privilege to
read from, write to, and deallocate the socket.  The third field is a
flag that indicates whether the socket has been allocated to a
particular process and, if so, how it can be used.  Outside the
context of MPD, the usage flag need indicate only whether the socket
is free or allocated.  However, since the model is created
specifically for use with MPD, the flag field of the socket descriptor
structure is also used to denote how the socket is used by an MPD
process.  For example, when a ring of MPD processes is established,
sockets corresponding to the right-hand side of the connection are
processed differently from the sockets on the left.
Figure~\ref{fig:skt-example-a}  
shows two connected MPD processes. Figure~\ref{fig:skt-example-b} 
shows the corresponding state of an array of socket descriptors as
well as one unallocated socket.  

The following Unix socket primitives have been modeled in accordance
with their defined functionality\cite{stevens-unp1}: \texttt{connect},
\texttt{accept}, \texttt{close}, \texttt{read}, and \texttt{write}.
Other socket primitives, such as \texttt{socket}, \texttt{bind}, and
\texttt{listen}, are not explicitly modeled but are implied in the
model.  \texttt{Select} is used only in the top tier of the MPD
process.  We include a model of \texttt{select} in the socket library,
but its implementation is greatly influenced by the specific way in
which the function is used by MPD algorithms.  The modeled socket
functions, defined as \texttt{inline} \promela\ functions of the
same name, serve as an interface between the bottom and middle tiers of
the model.  Below is an excerpt of the socket library together with the
definitions of the channels and socket descriptor structure.
Definitions of self-explanatory functions and macros are omitted.

\begin{verbatim}
chan connection[CONN_MAX] = [QSZ] of {msg_type};

typedef conn_info_type {
  unsigned other_fd  : FD_BITS; 
  unsigned owner_pid : PROC_BITS; 
  unsigned use_flag  : FLAG_BITS;
  };

conn_info_type conn_info[CONN_MAX];

inline read(file_desc, message) 
{       
  connection[file_desc]?message;
}

inline write(file_desc, message) 
{
  connection[conn_info[file_desc].other_fd]!message;
  fd_select_check(conn_info[file_desc].other_fd)
}

inline close(file_desc)
{
   IF  /* other side has not been closed yet */
   :: (conn_info[file_desc].other_fd != INVALID_FD) ->
      set_other_side(conn_info[file_desc].other_fd,INVALID_FD);
      fd_select_check(conn_info[file_desc].other_fd)
   FI;
   deallocate_connection(file_desc)
}

inline connect(file_desc, lp)
{       
   allocate_connection(j);          /* server's connection */
   set_owner(j, lp);
   set_handler(j, AWAIT_ACCEPT);
   allocate_connection(file_desc);  /* client's connection */
   set_owner(file_desc, _pid);
   set_other_side(j, file_desc);    /* relate connections */
   set_other_side(file_desc, j);    /* to each other      */
   lp_select_check(lp)
}

inline accept(file_desc)
{
   file_desc = 0;
   do
   :: (file_desc >= CONN_MAX) ->
      assert(0) /* block if no connect was made */
   :: (readable_lp(file_desc,_pid)) ->
      set_handler(file_desc, NEW);
      break
   :: else ->
      file_desc = file_desc + 1
   od
}
\end{verbatim}

Notice that \texttt{read(fd)} and \texttt{write(fd)} operate on
different sockets.  The reading operation amounts to receiving a
message from the \texttt{connection[fd]} channel, while the writing
operation places the message on the other buffer of the connection,
which is pointed to by the \texttt{other\_fd} field of
\texttt{conn\_info[fd]}.  This approach ensures that the connection is
truly bidirectional and that the two processes at the endpoints of the 
connection can read and write independently.    \texttt{Connect} and
\texttt{accept} are companion operations.  To establish a connection,
a client process connects to a listening port on the server process.
To complete the connection, the server must accept the connection
request from the client.  In the model, the client allocates both
sockets of the connection and sets the \texttt{use\_flag} of the
server's socket to \texttt{AWAIT\_ACCEPT}.   The \texttt{accept}
operation then locates the socket with the set flag.  In Unix, an
\texttt{accept} will block if executed before a \texttt{connect} is
issued.  The \promela\ model also implements this behavior.  Finally,
the \texttt{close} operation deallocates the socket.  When a
\texttt{close} is issued in Unix, the process on the other side of the
connection essentially sees an \texttt{EOF} on the corresponding
socket.  To simulate the behavior in \promela, the model of the
\texttt{close} operation sets the \texttt{other\_fd} of the remaining
part of the connection to an invalid value. \texttt{Write},
\texttt{connect}, and \texttt{close} operations contain references to
\texttt{\{fd,lp\}\_select\_check} functions, which are helper
functions to the \texttt{select} operation.  They set a bit for the
owner process of the connection, indicating that the process can leave
the idle state and do some processing in response to received messages
or other set flags.  The \texttt{select} function blocks until the bit
is set. 

\subsection{The Road Not Taken}

One may model the MPD system in many ways.  We are fully
prepared for the possibility that as this project progresses and we
model other kinds of MPD algorithms, our model will change.   The
three-layered approach allows us to make dramatic changes in different
levels of the model without affecting the remainder of the model.  We
experimented with several different models of the MPD system and came
to the following conclusions. 

While it is tempting to model the operating system explicitly and to
hide the manipulation of the sockets inside such a model, doing so
goes against the methodology of \spin/\promela\ model construction.  An
explicit model of the operating system produces a process that
simply forwards messages between daemons, managers, clients, and
consoles, resulting in a rapid explosion of the state space.

One can model the  socket operations by sending explicit
messages. For example, a \texttt{connect} can result in a message that
is consumed by the corresponding \texttt{accept}.  A \texttt{close}
operation can send a special \texttt{eof} message on the other end of
the connection, if it is still open.  The \texttt{select} operation
may be viewed as a message whose parameters include references to
sockets, which should be sampled for messages.  The \texttt{select}
would return when a response message is received.  The parameters of the
response message could contain the references to the sockets that in
fact have messages that are ready for consumption.  After
experimenting with such an approach, we decided that the model should
contain only those messages that represent explicit communication
between MPD processes, while communication that occurs between the
host operating systems should be represented by other means, such as
setting and resetting of flags. 


\section{Early Experiences with Modeling and Verification}
\label{sec:experiences}

This is our first \spin\ project.   Initial efforts concentrated on
investigating applicability of the \spin-based approach to our
problem.  We experimented by constructing different models of Unix
sockets and different models of MPD algorithms and attempting
verification of these models.  These efforts resulted in an
early success in demonstrating that a proposed MPD algorithm
was incorrect.\footnote{Admittedly, our focus at the time was on
  learning \spin\ and \promela\ and not on development of MPD
  algorithms.  Had we been concentrating on the latter topic, the
  rather obvious error we discovered with \spin's aid probably would
  have been found during early stages of design.  Nonetheless, we
  view the experience as evidence of the usefulness of the
  \spin-based approach to verification of MPD algorithms.}

\subsection{A Buggy Algorithm for Creating a Daemon Ring}

\begin{figure}[t]
    \centerline{ \epsfxsize=4.75in \epsfbox{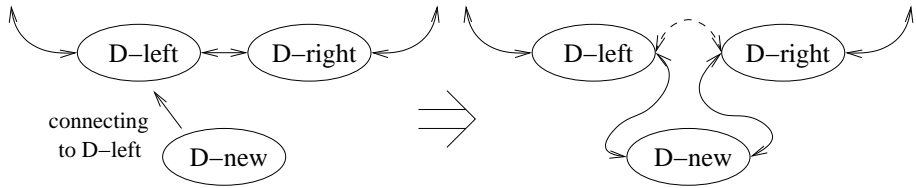} }
    \caption{Ring insertion}
    \label{fig:insert}
\end{figure}

Establishment of a ring of daemons, the first step of the MPD
system, and maintenance of the ring are central to the operation of
MPD.   Informally, daemon ring creation proceeds as follows.  The
initial daemon establishes a listening port to which subsequent
connections are made.  The daemon connects to its own listening
port, creating a ring of one daemon.  The listening port of the
first daemon and the name of the host processor are queried from the
console.  The desired number of daemons is then initiated and directed 
to enter the ring by connecting to the first daemon.
Figure~\ref{fig:insert}   
shows the result of inserting a new daemon into an existing ring.
Upon completion of the insertion the old connection between
daemons on the right and left of the new daemon is disconnected (shown
in the figure by the dashed line).  Note that in the special case of
insertion into a ring of one daemon, the daemon plays both the left
and the right roles.  

\begin{figure}[t]
    \centerline{ \epsfxsize=4.5in \epsfbox{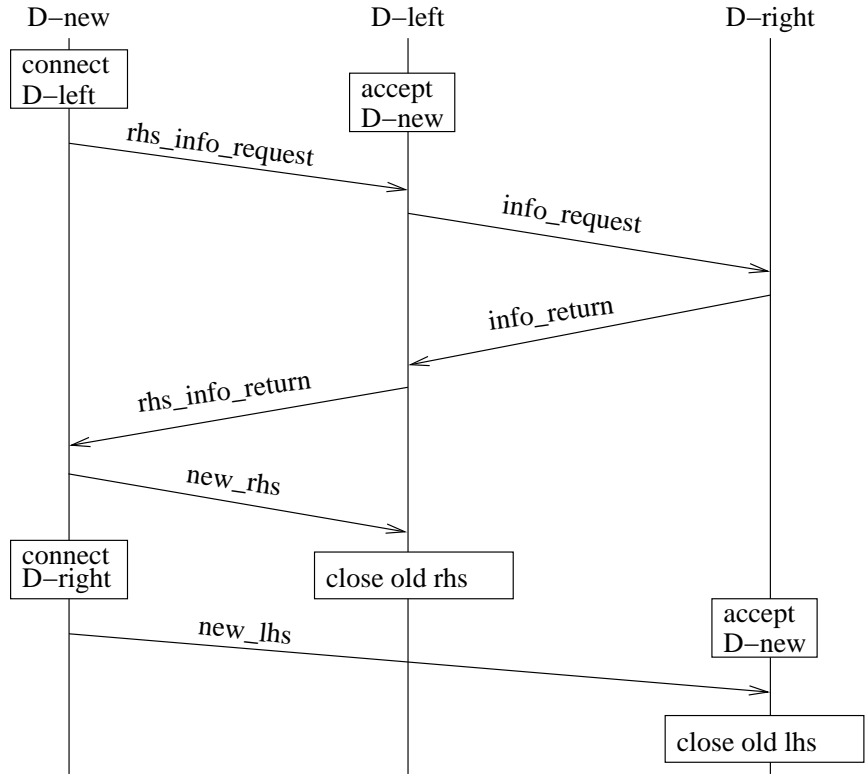} }
    \caption{Sequential ring insertion algorithm}
    \label{fig:insert-msc}
\end{figure}

Initially we modeled an algorithm that allowed new daemons to  enter
the ring sequentially.  Figure~\ref{fig:insert-msc}   
presents the algorithm as a message sequence chart.  The new daemon is
supplied with the information it needs to initiate a connection to the
daemon \textit{D-left}, so identified because it will end up as the
left-hand side neighbor of the new daemon.  The connection is accepted
by \textit{D-left}.   The new daemon then queries the \textit{D-left}
daemon about the coordinates of its right-hand side neighbor.  As no
information about the state of the ring is kept by the daemons in
this algorithm, \textit{D-left} queries \textit{D-right} for its
listening port number and the name of the host.  \textit{D-right}
returns the requested information, which is then forwarded to the new
daemon.  The new daemon declares itself the new right-hand side
neighbor of \textit{D-left}, which replaces its existing right-hand
side connection with the connection to \textit{D-new}.  The old
connection is closed.  The new daemon continues to enter the ring by
connecting to \textit{D-right} using the coordinates it received in
the \texttt{rhs\_info\_return} message from \textit{D-left}.  The
connection is accepted, and the new daemon declares itself the new
left-hand side of \textit{D-right}, which closes the remaining endpoint
of the old connection.

MPD design requirements, however, mandate that several ring insertion
commands can be issued simultaneously.  Moreover, an MPD process can
be blocked only by a select.  (The second requirement is fulfilled if
there are no blocking read statements within the body of any handler.)
Our sequential algorithm exhibits neither of the mandated properties.
The algorithm assumes that the sequence of messages
\texttt{rhs\_info\_request}, \ldots, \texttt{new\_rhs} is not 
interrupted, that is, \textit{D-new} sends an
\texttt{rhs\_info\_request} message and blocks until it receives the
\texttt{rhs\_info\_return} message. \textit{D-left} is similarly
blocked upon sending the \texttt{rhs\_info\_return} message until the
receipt of the \texttt{new\_rhs} message.   If the algorithm is
changed so that daemons are blocked only by a select and several new
daemons simultaneously enter the ring, an erroneous execution scenario
is possible.  The error occurs when two daemons attempt to
simultaneously enter a ring of more than one daemon.  The initial
connection is made by both new daemons to the same daemon in the
ring.  Then both daemons issue an \texttt{rhs\_info\_request}. If the
processing of the resulting messages is interleaved, both new
processes are told to connect their right-hand side to the same
daemon.  A ring will not be established in this case.  In the correct
algorithm, one of the new processes should connect its right-hand side
to the other new daemon, which should connect its right-hand side to
the existing daemon.   The algorithm is fixed by storing on each
daemon the coordinates of its right-hand side neighbor.  Doing so
makes unnecessary the series of messages to request and return this
information to the new daemon.

This error was discovered during simulations of the \promela\ model
within \spin.  The experience was important not only because \spin\
helped us find an error but also because \spin\ helped us to do so very
quickly.  This experience leads us to believe that the tool will become
invaluable in future development efforts both for verification and for
rapid prototyping and testing of new algorithms.  

\subsection{Early Problems with Modeling and Verification}

Our early models were naive because of our inexperience and our
attempts to match the model too closely to C code.  Including an
explicit model of the operating system for socket handling led to
extreme state space explosion while requiring a long state vector
size.  After examining the literature on pragmatic use of \spin\
\cite{spin:abstraction,spin:low-fat}, we developed much leaner models.
In optimizing our original explicit models we tried to find a level of 
abstraction that preserved the correlation of the \promela\ model to
the eventual C code while keeping verification feasible.   

\section{Verification of MPD Algorithms}
\label{sec:results}

An MPD daemon ring is a dynamic structure: new processes may
enter the ring at any time, and existing processes or their host
processors may crash or may shut down in an orderly fashion.  We
modeled and verified algorithms for ring creation and ring recovery
after a single process/processor crash. The majority of MPD algorithms
reside in the managers.  We modeled and verified a barrier algorithm,
an example of a manager-level functionality, which ensures that all
clients reach a certain point in the execution of a parallel job
before any client is allowed to proceed further. 

All verification runs were conducted on a 933 MHz Pentium III
processor with 970 MB of usable RAM.  We used default {\sc X}\spin\
settings for all verification attempts, except when we increased the
memory limit to allow the search to complete. In cases where
verification did not complete with default parameters within physical
memory limits, verification with compression (\texttt{-DCOLLAPSE}
compile-time directive) was performed.  Such experiments are
identified by an asterisk.   

\subsection{Ring Establishment Algorithm}

\begin{figure}[ht]
    \centerline{ \epsfxsize=4.5in \epsfbox{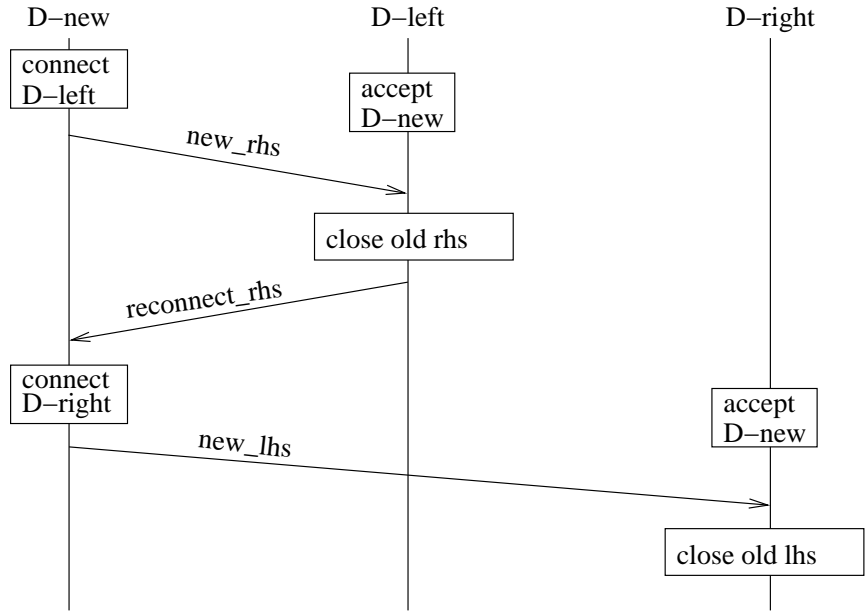} }
    \caption{Parallel ring insertion algorithm}
    \label{fig:insert-true-mpd}
\end{figure}

Figure~\ref{fig:insert-true-mpd} shows a message-sequence chart
representation of an algorithm that allows parallel insertion of
daemons into the ring.  A minor difference between this algorithm and 
the ring establishment algorithm described in
Section~\ref{sec:experiences} (Figure~\ref{fig:insert-msc}) is that
the \texttt{new\_rhs} message is sent before the coordinates of
\textit{D-right} are supplied to \texttt{D-new} via the message
\texttt{reconnect\_rhs}.  The major difference is that in this
algorithm each daemon records and maintains the identities (listening
port number and the host name) of the two daemons located to its
right.  Strictly speaking, only the identity of the immediate
right-hand side neighbor is needed to establish the ring.  To recover
from an unexpected crash of a daemon, however, the identity of the
second  right-hand side neighbor (rhs2) must be known as well.  The
algorithm involves sending a message of type \texttt{rhs2info}, to
communicate this information to the appropriate daemon.  This
operation is not shown in the figure.  In a small ring (fewer than
three daemons) or when a ring is being created, special case logic
ensures that appropriate information is forwarded to the correct
daemon.  In the general case, upon receipt of the \texttt{new\_rhs}
message \textit{D-left} sends the \texttt{rhs2info} message, in a
counterclockwise direction along the ring, to its left-hand side
neighbor,  indicating that it has a new rhs2, namely, \texttt{D-new}.
In addition, upon receipt of the  \texttt{new\_lhs} message,
\textit{D-right} sends the \texttt{rhs2info} message along the ring,
also in a counterclockwise direction, to \textit{D-new}.   The rhs2
neighbor of  \textit{D-new} is the right-hand side neighbor of
\textit{D-right}.  

The algorithm was modeled by using the three-tiered approach described
above.  To check whether the algorithm is correct, we verify that the
resulting system topology, which is implicit in the socket descriptor
structures array, is in fact a ring of the correct size.  In addition,
we check that the state information (identities of the two neighbors
to the right) agrees with the information in the socket descriptor
structures array.  These two checks are performed when the ring
establishment algorithm is completed, which in reality corresponds to
all daemons entering an idle state.  In the \promela\ model this
corresponds to a timeout.   A properly connected ring manifests itself  
in successfully supporting operation of daemon-level algorithms.  In
fact, MPD designers and users test whether the ring was established
successfully by executing one of such algorithms and examining its
results.  The algorithm, invoked by \texttt{mpitrace}, reports the
identities of all daemons and their position in the ring.  To convince
our ``customers'' (i.e., MPD designers) that the \promela\ model of
the ring establishment algorithm is correct, we also verify that upon
its completion, the trace algorithm terminates after having visited
every daemon in the ring.   The trace algorithm is implemented both in
MPD and in our model by sending two kinds of messages along the ring,
which leads to an increase in the number of states that have to be
examined during a verification attempt.  For this reason, on a model
of a given size, verification of the successful trace completion is
more expensive, in terms of time and memory, than verification of ring
topology and state information.

\begin{table}[t]
\centerline{
\begin{tabular}{c|c|c|c|c|c|c} 
Correctness & Model & Time & Memory & Vector Size & States & Search  \\ 
Property    & Size  & (s)  & (MB)   & (byte)      &Stored/Matched &
Depth \\ \hline
State & 1  & 0.00   & 2.5 &  40 & 10/0              & 9 \\
      & 2  & 0.00   & 2.5 &  92 & 44/23             & 21\\
      & 3  & 0.05   & 3.0 & 136 & 4304/5278         & 44 \\
      & 4  & 105.35 & 768 & 224 & 3.83e+06/7.97e+06 & 115 \\ \hline
Trace & 1  & 0.00   & 2.5 & 40  & 14/0              & 13 \\ 
      & 2  & 0.00   & 2.5 & 92  & 56/27             & 29 \\
      & 3  & 0.80   & 3.3 & 136 & 5743/6718         & 58 \\
      & 4* & 159.33 & 173 & 224 & 4.57e+06/9.28e+06 & 115 \\ 
\end{tabular}}
\vspace{0.1in}
    \caption{Verification statistics for the ring establishment algorithm}
    \label{tbl:ring}
\end{table}
Statistics for the verification of the ring algorithm are presented in
Table~\ref{tbl:ring}.  The first portion of the table reports on the
verification of state of the ring properties, that is, the information
in the socket description structures array and the recorded identities
of the two right-hand side neighbor of each daemon.  The second
portion of the table presents statistics on verification of a
successful trace after ring establishment.  The algorithm was verified
for models comprising up to four daemons, as indicated in the second
column of the table. 

We were unable to exhaustively verify the algorithm on models with
five or more daemons. Because of dramatic state space explosion and
large state vectors,  verification attempts for these models ran out
of 970 MB of memory, and using the compression and graph encoding
techniques (\texttt{-DCOLLAPSE} and \texttt{-DMA=n} compile-time
directives) still did not enable the search to complete.  When we
applied predicate abstraction techniques to model the ring
establishment algorithm, verification succeeded for models with only
up to eight daemons.  However, the desired correlation of the
\promela\ model to the C/Python code was lost, as was the ability to
perform meaningful simulations.  As shown below in verification
statistics of other MPD algorithms that we have modeled and verified,
the rapid increase in the number of states for relatively small models
occurs only for the ring establishment algorithm and is due in large
part to its properties.  

Since the daemons enter the ring in parallel, there are many possible
interleavings  of their execution leading to many possible ring
configurations. In general, given $n$ daemons, there are $n!$
resulting ring configurations.  Moreover, a daemon enters the ring in  
two steps: first it connects its left-hand side, then its right-hand
side, which in turn increases the number of possible interleavings.
An increase in model size also leads to an increase in the number of
sockets/fds that are manipulated.  Since every connection consists of
two sockets, there is a minimum of $2n$ sockets in a ring of $n$
daemons.  However, as the ring is being created, additional sockets
are required because there are execution sequences in which allocation
of a new socket occurs {\em before} deallocation of a socket belonging
to a connection whose other end has been closed.  Finally, an increase
in model size leads to an increase of the buffer size of the
communication channels.  Recall that when a new daemon enters the
ring, a \texttt{rhs2info} message is sent along the ring in a
counterclockwise direction to a daemon located two positions to the
left of the new daemon.  Therefore, in a general case, for $n$ new
daemons inserted into the ring, $n$ \texttt{rhs2info} messages are
sent.  Some execution sequences result in all $n$ messages
accumulating on a single communication channels.  Therefore, the
buffer size is equal to the number of daemons. 

The fact that verification succeeds only on small models needs to be
put in perspective.  On one hand, a running MPD typically consists of
several tens or hundred of processes.  Therefore, given the current
technology, we cannot completely verify MPD algorithms for every
possible system size and topology. On the other hand, prior experience
with debugging of the MPD code suggests that even the most difficult
errors manifest themselves in systems of just a few (four to ten)
processes.  Therefore, the models of MPD algorithms as the current
level of abstraction allow us to perform verification of some
algorithms on models of satisfactory size.  For other algorithms, such
as the ring establishment algorithm, a slightly more abstract model or
a more efficient socket library will enable verification to complete
on models of sufficient size.

\subsection{Recovery from a Single Nondeterministic Failure}

The ring recovery algorithm works as follows.  When a daemon in a
properly established ring fails, the operating system on its host
processor will close all sockets that belonged to it, so the neighbors
of the failed daemon will see that the sockets on their opposite end
of their connection to the daemon were closed.  Of course, sockets may
be closed for legitimate reasons as well, but the suite of MPD
algorithms is designed in such a way that a closed socket on the
right-hand side connection in the ring, without any advance
notification that a controlled shutdown or disconnection will take
place, signifies an unintended failure and places the burden of
recovery on the remaining members of the ring.  In the algorithm, the
left-hand side neighbor reinstates the ring by establishing a connection
to the right-hand side neighbor of the failed daemon.  The identities of
the two right-hand side neighbors must be updated for all affected
daemons.   

\begin{table}[ht]
\centerline{
\begin{tabular}{c|c|c|c|c|c|c} 
Correctness & Model & Time & Memory & Vector Size & States & Search  \\ 
Property    & Size  & (s)  & (MB)   & (byte)      &Stored/Matched &
Depth \\ \hline
State & 2   & 0.00    & 2.5   & 96  & 52/2                    & 34 \\
      & 3   & 0.00    & 2.5   & 116 & 266/141                 & 56 \\
      & 4   & 0.01    & 2.6   & 180 & 1073/1288               & 73\\
      & 5   & 0.10    & 3.4   & 216 & 4276/7985               & 90 \\
      & 6   & 0.60    & 7.0   & 304 & 16323/41086             & 107\\
      & 7   & 2.97    & 23.1  & 352 & 59822/189011            & 124 \\
      & 8   & 15.79   & 91.9  & 464 & 212205/806732           & 141 \\
      & 9   & 69.70   & 376.0 & 520 & 734040/3.26e+06   & 158 \\
      & 10* &  243.08 & 71.8  & 656 & 2.49e+06/1.27e+07 & 175 \\ 
      & 11* & 1054.09 & 235.9 & 720 & 8.32e+06/4.78e+07 & 192 \\ 
      & 12* & 5340.19 & 772.6 & 876 & 2.75e+07/1.76e+08 & 209 \\ 
\hline 
Trace & 2   & 0.00    & 2.5   & 96  & 64/2                   & 38 \\
      & 3   & 0.01    & 2.5   & 116 & 326/161                & 64 \\
      & 4   & 0.02    & 2.7   & 176 & 1433/1648              & 87 \\
      & 5   & 0.17    & 4.0   & 216 & 6992/12801             & 112\\
      & 6   & 1.46    & 13.1  & 304 & 38755/98302            & 139 \\
      & 7   & 13.74   & 89.7  & 352 & 253322/832511          & 168 \\
      & 8   & 163.39  & 814.1 & 464 & 1.93e+06/7.83e+06 & 199 \\
      & 9*  & 1919.94 & 502.5 & 520 & 1.62e+07/7.85e+07 & 232 \\
\end{tabular}}
\vspace{0.1in}
    \caption{Verification statistics for the ring recovery algorithm}
    \label{tbl:crash}
\end{table}

In our model of the algorithm the initial ring is hard-coded.  One
daemon is directed to fail, although we do not specify which one
should do so.  The recovery procedure is then initiated.  The model
algorithm was verified against the three correctness properties of a
properly established ring, as discussed above.  Table~\ref{tbl:crash} 
shows statistics of the verification attempts. 

\subsection{Manager-Level Barrier Algorithm}

Parallel programs frequently rely on a \textit{barrier} mechanism to
ensure that all processes of the job reach a certain point (complete
initialization, for example) before any are allowed to proceed
further.  Parallel jobs, that is, programs running on the clients, rely
on the manager processes to implement the barrier service.  The
algorithm proceeds as follows.  A manager process is designated as
the leader of the algorithm and is given a rank of 0.   When the
leader reads a request from its client to provide the barrier service,
it sends a message \texttt{barrier\_in} to its right-hand side neighbor
in the ring.  When a non-leader manager receives the
\texttt{barrier\_in} message, its behavior is determined by whether its
client has already requested the barrier service.  If the client has
done so, the manager forwards the message to the right-hand side
manager.  Otherwise, it holds the \texttt{barrier\_in} message until
the request from the client arrives.  While the \texttt{barrier\_in}
message is held, a bit variable \texttt{holding\_barrier\_in} is set.
Once the \texttt{barrier\_in} message traverses the entire manager ring
and arrives back in the leader, meaning that each client has reached
the barrier and notified its manager, the leader sends a
\texttt{barrier\_out} message around the ring.  When a manager receives
the \texttt{barrier\_out} message, it notifies its client to proceed
past the barrier.  The leader can be either the first or the last
manager to allow its client to proceed.

We modeled the algorithm on top of the socket library.  A ring of
managers is hard-coded in the beginning, and a manager with
\texttt{\_pid} of 0 is designated as the leader.  There is no need to
model the clients explicitly; thus the communication with them is
represented by two bits per manager.  One bit designates that a request
for a barrier service has been received by the manager.  The other bit,
\texttt{client\_barrier\_out}, designates that the manager notified the
client to proceed past the barrier.  Global arrays of bits were
defined, using the bit-array implementation by Ruys
\cite{spin:low-fat}, to store these values.  A constant 
\texttt{ALL\_BITS} corresponds to a value of the bit array where an
element was set for every manager in the ring.

We verified two correctness conditions about the algorithm.  First, at
the end of the algorithm, all clients must have been told to proceed
past the barrier:

\begin{verbatim}
timeout -> assert(client_barrier_out == ALL_BITS)
\end{verbatim}

\noindent
The second condition is an invariant:  no client is allowed to proceed
until all clients have reached the barrier and all managers have
released the \texttt{barrier\_in} message: 

\begin{verbatim}
assert((client_barrier_out == 0) || 
       ((client_barrier_in == ALL_BITS) 
        && (holding_barrier_in == 0)))
\end{verbatim}

Table~\ref{tbl:barrier} shows statistics for verification of the
barrier algorithm.  We were able to exhaustively verify models with
up to fourteen managers. 

\begin{table}[t]
\centerline{
\begin{tabular}{c|c|c|c|c|c} 
 Model & Time & Memory & Vector Size & States & Search  \\ 
 Size  & (s)  & (MB)   & (byte)      &Stored/Matched & Depth \\ \hline
1   & 0.00   & 2.5   & 40  & 20/4              & 14 \\
2   & 0.00   & 2.5   & 60  & 47/24             & 21 \\
3   & 0.00   & 2.5   & 84  & 118/118           & 29 \\
4   & 0.00   & 2.5   & 108 & 321/506           & 37 \\
5   & 0.01   & 2.6   & 128 & 920/1992          & 45 \\
6   & 0.04   & 2.8   & 152 & 2707/7420         & 53 \\
7   & 0.18   & 3.8   & 172 & 8058/26618        & 61 \\
8   & 0.73   & 7.2   & 196 & 24101/92958       & 69 \\ 
9   & 2.42   & 18.1  & 220 & 72220/318220      & 77 \\
10  & 9.73   & 54.6  & 244 & 216567/1.07e+06   & 85 \\
11  & 35.03  & 168.8 & 264 & 649598/3.57e+06   & 93 \\
12  & 127.97 & 549.2 & 288 & 1.95e+06/1.18e+07 & 101\\
13* & 759.23 & 192.9 & 308 & 5.85e+06/3.85e+07 & 109 \\
14* & 3050.96 & 571.3 & 332 & 1.75e+07/1.25e+08 & 117
\end{tabular}}
\vspace{0.1in}
    \caption{Verification statistics for the barrier algorithm}
    \label{tbl:barrier}
\end{table}

\section{Summary and Future Plans}
\label{sec:plans}

We described here our first experiences in applying the \spin-based
approach to verification of a parallel process management system called
MPD.  We settled on a three-tier architecture for the models in order
to maintain the correlation to the eventual code of the MPD system and
to enforce modularity of the model.  The bottom layer of the
architecture consists of a \promela\ model of operations on Unix
sockets.  We encountered some early difficulties in the verification
attempts.   Specifically, for the ring establishment algorithm, exhaustive
verification can be completed only on models with up to four daemons.
However, we were able to exhaustively verify larger models of other
algorithms. 

Based on our experiences, we believe that design and development of
algorithms for MPD and similar systems can  benefit greatly from
application of the \spin-based software verification methods. \spin's
simulation capability allows for rapid prototyping of new algorithms.
Since even the most difficult errors can be discovered on models
comprising only a few processes, the verification engine of \spin\
enables us to verify the algorithms on models that are sufficiently
large for our purposes. 

A long-term goal of this project is to model and verify MPD algorithms
and then translate them into C or another programming language, while
preserving the verified properties.  Ideally, translation should be
automated. To allow this to happen, the \promela\ model
must not be overly abstract.  Figure~\ref{fig:code} shows a \promela\
\begin{figure}[ht!]
\begin{verbatim}
p.1  :: (msg.cmd == barrier_in) ->
p.2     if
p.3     :: (IS_1(client_barrier_in,_pid)) ->
p.4        if
p.5        :: (_pid == 0) ->
p.6           make_barrier_out_msg;
p.7           find_right(fd,_pid);
p.8           write(fd,msg)
p.9        :: else ->
p.10          make_barrier_in_msg;
p.11          find_right(fd,_pid);
p.12          write(fd,msg)
p.13       fi
p.14    :: else ->
p.15       SET_1(holding_barrier_in,_pid)
p.16    fi

c.1  if ( strcmp( cmdval, "barrier_in" ) == 0 ) {
c.2     if ( client_barrier_in ) {
c.3        if ( rank == 0 ) {
c.4           sprintf( buf, 
                       "cmd=barrier_out dest=anyone src=%s\n", 
                       myid );      
c.5           write_line( buf, rhs_idx );
c.6        }
c.7        else {
c.8           sprintf( buf, 
                       "cmd=barrier_in dest=anyone src=%s\n", 
                       origin );    
c.9           write_line( buf, rhs_idx );
c.10       }
c.11    }
c.12    else {
c.13       holding_barrier_in = 1;
c.14    }
c.15 }
\end{verbatim}
    \caption{Portion of the \promela\ model and C implementation of
      the barrier algorithm}
    \label{fig:code}
\end{figure}
model and a C implementation of a portion of the barrier algorithm, in
which a \texttt{barrier\_in} message is received and processed by a
manager.  Automated translation certainly appears feasible for this
level of abstraction of the \promela\ model.  Notice the one-to-one
correspondence between the control structures of the two segments.
There is further correspondence between \promela\ and C for checking
and setting of boolean variables (lines \texttt{p.3},\texttt{c.2} and 
\texttt{p.15},\texttt{c.13}).   The code for message assembly (lines
\texttt{p.6},\texttt{c.4} and \texttt{p.10},\texttt{c.12}) matches
as follows. A \promela\ macro \texttt{make\_barrier\_out\_msg}
corresponds in C to writing a string, containing a
\texttt{barrier\_out} command, to a buffer.  In the \promela\ model,
before a message can be written to a buffer, the corresponding fd
must be found using functions \texttt{find\_right} or
\texttt{find\_left}.  In the C code, the buffer is written to a file
descriptor correspondingly referenced by \texttt{rhs\_idx} or
\texttt{lhs\_idx}. Therefore, the two \promela\ lines \texttt{p.6-7}
match a single C line \texttt{c.5}.  

Of course, not all models will lend themselves well to verification at
this level of abstraction, as verification of the ring establishment
algorithm demonstrated.  But, models of the MPD algorithms should fall
into just a few different classes with respect to the level of
abstraction, and a separate mapping can be defined for each such level
to enable the \promela-to-C translation. 

We will continue to model and verify individual MPD algorithms.  They
include the daemon-level functionality for controlled shutdown of a
portion of the ring and subsequent ring reestablishment, as well as
various manager-level algorithms, such as the handling of the parallel
input and output to the console.   Correct interaction of these
algorithms is also very important.  Many things take place in parallel
in a running MPD system.  Daemons enter and leave the ring, as do
managers, different client processes request different services
from the managers, and several instances of the same algorithm may be
executing simultaneously.  We hope to be able to reason formally about
MPD models that consist of several related and interdependent
algorithms.   

The \promela\ implementation of the the Unix socket library as well as
models of the MPD algorithms described in this paper are available at 
\texttt{http://www.mcs.anl.gov/\homedir matlin/spin-mpd}.


\bibliographystyle{plain}
\bibliography{spin.bib}



\end{document}